\documentclass[graybox]{svmult}

\usepackage{mathptmx}       
\usepackage{helvet}         
\usepackage{courier}        
\usepackage{type1cm}        
\usepackage{makeidx}         
\usepackage{graphicx}        
\usepackage{multicol}        
\usepackage[bottom]{footmisc}

\makeindex             

\begin{document}

\title*{Statistical Mechanics of Labor Markets}
\author{He Chen and Jun-ichi Inoue}
\institute{He Chen \at Graduate School of Information Science and Technology, 
Hokkaido University, N14-W-9, Kita-ku, Sapporo 060-0814, Japan \email{chen@complex.ist.hokudai.ac.jp}
\and Jun-ichi Inoue \at Graduate School of Information Science and Technology,
Hokkaido University, N14-W-9, Kita-ku, Sapporo 060-0814, Japan \email{jinoue@cb4.so-net.ne.jp}}
%
%
\maketitle

\abstract{
We introduce a probabilistic model of labor markets for university graduates, in particular, in Japan. 
To make a model of the market efficiently, we take into account several hypotheses. 
Namely, each company fixes the (business year independent) number of opening positions for newcomers. 
The ability of gathering newcomers depends on the result of job matching process in past business years. 
This fact means that the ability of the company is weaken if the company did not make their quota or the company gathered applicants too much over the quota. 
All university graduates who are looking for their jobs can access the public information about the ranking of companies. Assuming the above essential key points, we construct the local energy function of each company and describe the probability that an arbitrary company gets students at each business year by a Boltzmann-Gibbs distribution. 
We evaluate the relevant physical quantities such as the employment rate. 
We find that the system undergoes a sort of `phase transition' from the `good employment phase' to `poor employment phase' when one controls the degree of importance for the ranking.  
}
\section{Introduction}
Deterioration of the employment rate is now one of the most serious problems in Japan and various attempts to overcome these difficulties have been done by central or local governments. 
Apparently, labor (work) is important not only for each of us to earn our daily bread, but also for our state to keep the revenues by collecting the taxes from labors. 
Especially, in recent Japan, the employment rate is getting worse and the government has over-issued quite a lot of government bonds to compensate a lack of the tax revenues and the national debt (amount to $\sim 6 \times 10^{15}$ Japanese yen or more!) is now becoming a serious risk to cause a national-wide bankruptcy. 

To make the matter worse, the earthquake and tsunami hit the northeast coast on 11th March 2011, and as the result, 
Fukushima nuclear power plant was seriously damaged and people living in that area has taken refuge from the nuclear radiation. 
These unpredictable disasters caused by nature and human error have made our country in financial difficulties that we have never encountered before. 
Many people lost their jobs and a lot of companies and plants could not be maintained. 
Hence, the reconstruction (improvement) of working condition for the labors and companies is now not something that can be ignored.

To consider the effective policy and to carry out it for sweeping away the unemployment uncertainty, it seems that we should investigate the labor markets scientifically and if it is possible, one should simulate artificial labor markets in personal computer to reveal the essential features of the problem. 
In fact, in macroeconomics (labor science), there exist a lot of effective attempts to discuss the macroscopic properties \cite{Aoki,Boeri,Roberto,Fagiolo,Casares,Neugart}. 
However, apparently, the macroscopic approaches lack of their microscopic view point, namely, in their arguments, the behaviour of microscopic agents such as job seekers or companies are neglected.

Taking this fact in mind, in this paper, we shall propose a simple probabilistic model based on the concept of statistical mechanics for stochastic labor markets, in particular, Japanese labor markets for university (college) graduates. 
\section{Empirical evidence: The Philips curve}
Let us first mention the relationship between the unemployment 
and the inflation rates in recent Japan (from 1970s to 2000s) in empirical evidence of the labor market. 
This relationship is generally called as {\it Philips curve}. 
In the original paper of Philips \cite{Philips}, 
he found the relation for the empirical data set from the middle of nineteen century to the beginning of twenty century (1861-1913) in UK. 
However, up to now, a lot of verifications have been done for various data sets in various countries. 
Therefore, we are confirmed that the Philips curve should be regarded as one of the `universal properties' in labor markets. 

\begin{figure}[ht]
\begin{center}
\includegraphics[width=5.5cm]{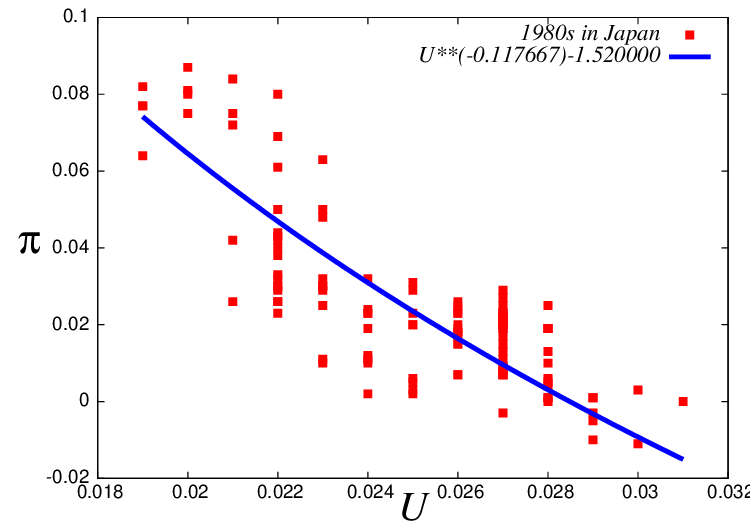} 
\includegraphics[width=5.5cm]{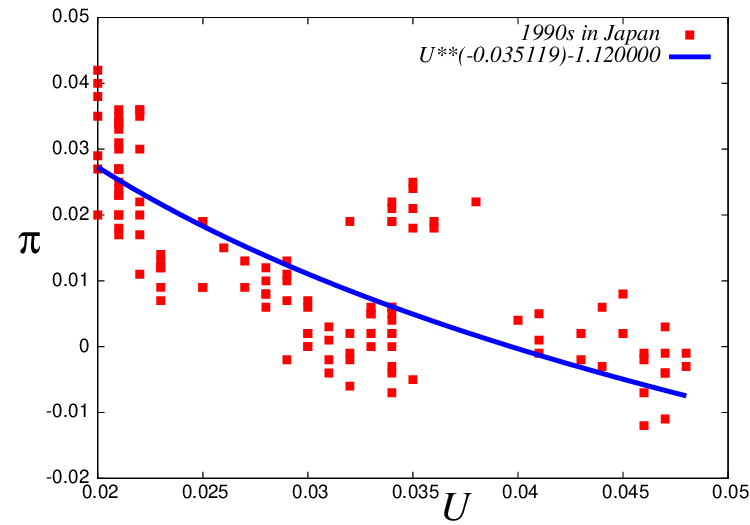} \\
\includegraphics[width=5.5cm]{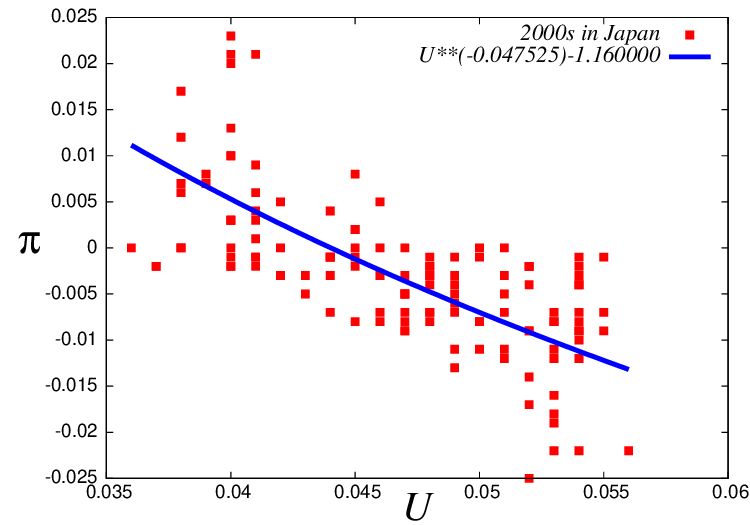}
\includegraphics[width=5.5cm]{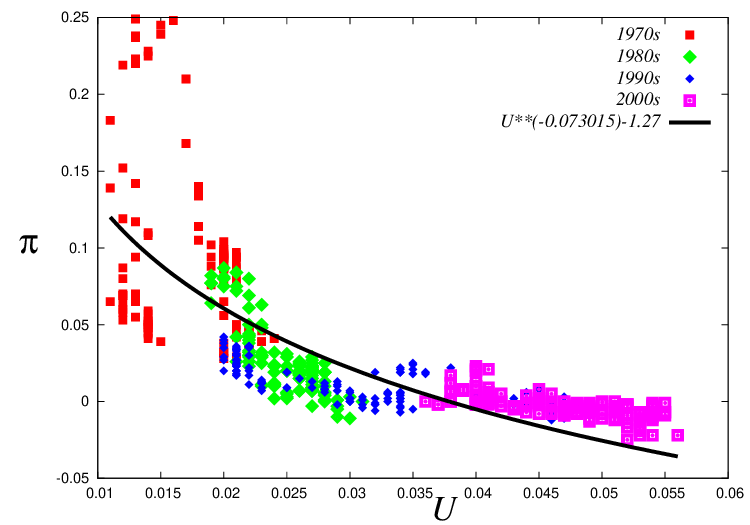}
\end{center}
\caption{\footnotesize 
The Philips curves in Japan. From the upper left to the lower right, the curves in 1980s, 90s and 2000s are plotted. In the lower-right panel, 
Philips curve in Japan (from 1970s to 2000s) is shown by a single plot. 
We clearly find that the Philips curve is getting `flat'. 
}
\label{fig:fg03}
\end{figure}

In Fig. \ref{fig:fg03}, we plot the Philips curves of our country in 1980s, 90s and 2000s and the age-dependence of the curve (the lower-right panel). 
In these panel, the fitting curve (the solid curve): 
$\pi +b \propto U^{-c}$  is actually obtained by the least square estimation for the parameter $b$ and $c$ from the $n$-data points $(X_{i},Y_{i})=
(\log U_{i}, \log (\pi_{i}+b)),\,\,i=1,\cdots,n$.  
From the lower-right panel in Fig. \ref{fig:fg03}, we clearly find that the curve is slowly getting `flat'.

From these empirical findings, we easily notice that the curve changes (evolves) in time and the behaviour might be dependent on the situation of economy in the country. 
Therefore, it is important for us to reveal the dependence from the microscopic point of view and it is our motivation to deal with the problem in this paper. 
\section{Hypotheses in modeling}
We first discuss several basic properties should be satisfied in our probabilistic modeling. 
As we already mentioned, in our modeling, the job seekers are restricted to university graduates and the other persons on the job searching are neglected for simplicity. 
Strictly speaking, this assumption should not be justified, however, as is well known, 
Japanese society is still somewhat conservative and once labors get their jobs, they stay in the company by their retirement age. 
In this sense, we omit the contribution from the on the job searching persons to the labor market because the fraction at the present time might be negligibly small.  

Based on the above general `hypotheses', we assume that the following three points (i)-(iii) should be taken into account to construct the effective labor markets. 
\begin{enumerate}
\item[(i)]
Each company recruits constant numbers of newcomers in each business year.  
\item[(ii)] 
If the company takes too much or too less applications which are far beyond or far below the quota,
 the ability of the company to gather applicants in the next business year decreases. 
\item[(iii)]
Each company is apparently ranked according to various perspectives such as the stability, 
the wage-level, the location, the welfare of employees, {\it etc}. The ranking is useful information and it is available for all students. 
\end{enumerate}
In following, we shall attempt to construct labor markets by considering the above three essential points. 
Our model system is a variant of urn models \cite{Aoki,Enrico} or the so-called {\it Kolaka Paise Restaurant Problem (KPRP)}\cite{Bikas, Asim}. 
\section{System parameters}
First of all, in this section, we define the system parameters and their sizes such as the number of companies, open positions (vacancies) and applicants. 
Let us define the total number of companies as $K$ and each of them is distinguished by the label $k=1,2, \cdots,K$. 
Then, the number of the quota of the company $k$ is specified by $v_{k}^{*}$. 
In real labor markets, the quota $v_{k}^{*}$ itself fluctuates in time (business year) and it changes according to the gross margin in the previous year and some companies in financial difficulties might decrease the quota. 
However, in this paper, we fix the value and regard the quota as a `uniform' and `time-independent' variable.
 
Hence, the total job vacancy in society in each business year $V$ is given by 
\begin{equation}
V  =  \sum_{k=1}^{K}v_{k}^{*}. 
\end{equation}
On the other hand, we define the number of new university graduates by $N$ and each of the students is specified by the index $i$ as 
$i=1,2,\cdots, N$. 
Then, we introduce 
\begin{equation}
\alpha  \equiv   \frac{V}{N}
\end{equation}
as {\it job offer ratio} and it is independent of system size for $\mathcal{O}(V)=\mathcal{O}(N)$. 
Apparently, for $\alpha  =V/N >1$, that is 
$V>N$, the labor market behaves as a `seller's market', whereas for $\alpha < 1$, the market becomes a `buyer's market'. 
For this model system, we might assume that each student post his/her single application (what we call `entry sheet' or CV) to the company. 
In our analysis given below, 
the relevant system parameter is job offer ratio $\alpha$ rather than $V$ or $N$ because obviously the $\alpha=V/N$ is system size independent. 
\section{The local energy function: A link to physics}
Here we define a sort of `local energy function' for each company which represents the ability of gathering applicants in each business year $t$. 
The energy function is a nice bridge to link the labor market to physics. 
Let us first define the following {\it local mismatch measurement}: $h_{k}(t)$ for each company $k\, (=1,2,\cdots,K)$ as 
\begin{equation}
h_{k}(t) =  
\frac{1}{V} 
|v_{k}^{*}-v_{k}(t)| = 
\frac{1}{\alpha N} 
|v_{k}^{*}-v_{k}(t)|
\end{equation}
where $v_{k}(t)$ denotes the number of students who seek for the position in the company $k$ at the business year $t$ (they will post their own `entry sheet (CV)' to the company $k$). 
Hence, the local mismatch measurement $h_{k}(t)$ is the difference between the number of applicants $v_{k}(t)$ and the quota $v_{k}^{*}$. 
We should keep in mind that from the fact (i) mentioned before, the $v_{k}^{*}$ is a business year $t$-independent constant.  
Some analysts reported that the increase of unemployment rate in recent Japanese labor market is due to the mismatch between the university graduates and companies. 
Namely, most of students look for the positions in famous and already established large companies and they do not want to work for a small business in, 
say, like fostering venture businesses. Such a sort of `local mismatch' could be quantified by $h_{k}(t)$. 

On the other hand, we define the ranking of the company $k$ by $\epsilon_{k} (>1)$ which is independent of the business year $t$. 
Here we assume that the ranking of the company $k$ is higher if the value of $\epsilon_{k}$ is larger. 
In this paper, we simply set the value as  
\begin{equation}
\epsilon_{k}  =  1+\frac{k}{K}
\label{eq:ranking}. 
\end{equation}
Namely, the company $k=K$ is the highest ranking company, whereas the company $k=1$ is the lowest. 

Form the above set-up and on the analogy of the Boltzmann-Gibbs distribution in conventional statistical mechanics, we define the probability 
$P_{k}(t)$ that the company $k$ gathers their applicants at time $t$ as 
\begin{equation}
P_{k}(t)   =  
\frac{\epsilon_{k}}{Z}\,
{\exp}
\left[
-H_{k}(\mbox{\boldmath $\beta$}_{k},\mbox{\boldmath $h$}_{k})
\right] \equiv     
\frac{
{\exp}
\left[
-E(\epsilon_{k}, H_{k}(\mbox{\boldmath $\beta$}_{k},\mbox{\boldmath $h$}_{k})
\right]}{Z}
\label{eq:prob} 
\end{equation}
\begin{equation}
Z \equiv 
\sum_{k=1}^{K}
{\exp}
\left[-E(\epsilon_{k}, H_{k}(\mbox{\boldmath $\beta$}_{k},\mbox{\boldmath $h$}_{k})
\right] 
\label{eq:norm}
\end{equation}
where we defined $Z$ as the normalization constant for probability (a sort of partition function in statistical physics). 
We also defined two $\tau$-dimensional {\it market history vectors} with the length $\tau$: 
$\mbox{\boldmath $\beta$}_{k} \equiv 
(\beta_{k}(t-1),\cdots,\beta_{k}(t-\tau))$ and 
$\mbox{\boldmath $h$}_{k} \equiv 
(h_{k}(t-1),\cdots,h_{k}(t-\tau))$. 
Then, $H_{k}$ appearing in the probability (\ref{eq:prob}) is defined by the inner product of these two vectors as 
\begin{equation}
H_{k}(\mbox{\boldmath $\beta$}_{k},\mbox{\boldmath $h$}_{k})
\equiv \mbox{\boldmath $\beta$}_{k} \cdot 
\mbox{\boldmath $h$}_{k}
\label{eq:inner}.
\end{equation}
We should notice that the above inner product choice for the expression of $H_{k}$ is just an example and one can easily extend (modify) the functional form to much more generalized one: 
$H_{k} = f(\mbox{\boldmath $\beta$}_{k},\mbox{\boldmath $h$}_{k})$ including (\ref{eq:inner}) as a special case. 

With the above definitions, the local energy function is written explicitly by
\begin{equation}
E(\epsilon_{k}, H_{k}(\mbox{\boldmath $\beta$}_{k}, \mbox{\boldmath $h$}_{k}))  \equiv  
-\gamma \log \epsilon_{k} + 
\sum_{l=1}^{\tau}
\beta_{l}h_{k}(t-l).
\label{eq:energy}
\end{equation}
In this paper, we simply choose a particular market history vector as 
$\mbox{\boldmath $\beta$}_{k} = (\beta,0,\cdots,0)$. 

Thus, the local energy function (\ref{eq:energy}) is now simplified as 
\begin{equation}
E(\epsilon_{k},h_{k}(t-1)) = 
-\gamma \log \epsilon_{k} + 
\beta h_{k}(t-1). 
\end{equation}
The parameters $\gamma >0$ and $\beta$ specify the probability from the macroscopic point of view. 
Namely, the company $k$ having relatively small 
$h_{k}(t)$ can gather a lot of applicants in the next business year and the ability is controlled by the parameter 
$\beta$ (we used the fact (ii) which was mentioned in the previous section). 
On the other hand, the high ranked company can gather lots of applicants and the degree of the ability is specified by the parameter $\gamma$  (we used the fact (iii) which was mentioned in the previous section). 

Thus, local energy function $E(\epsilon_{k},h_{k}(t-1))$ is written in terms of the sum of these two independent factors. 
Therefore, the result of the previous business year $h_{k}(t)$ is much more important factor for $\gamma > \beta$ and the ranking becomes more essential for $\gamma < \beta$ to decrease the energy. 

We should bear in mind that even if the highest ranking company $k=K$ gathers a lot of applicants as over the quota 
$v_{K}(t) \gg v_{K}^{*}$ at some year $t$, however, the second term appearing in the energy function 
$\sim -\beta |v_{K}^{*}-v_{K}(t)| \ll 1$ acts as the `negative feedback' on the first ranking preference term 
to decrease the probability that the company $K$ gathers the applicants at the next business year $t+1$. 
\section{Job matching process: microscopic quantities}
We should notice that for the probability $P_{k}(t)$, each student $i$ decides to post their entry sheet to the company $k$ at time $t$ as 
\begin{equation}
a_{ik}(t)  =  
\left\{
\begin{array}{cl}
1 & (\mbox{with prob. $P_{k}(t)$})\\
0 & (\mbox{with prob. $1-P_{k}(t)$}) 
\label{eq:aik}
\end{array}
\right.
\end{equation}
where $a_{ik}=1$ means that the labor $i\,(=1,\cdots,N)$ post their entry sheet to the company $k$ 
and $a_{ik}=0$ denotes that he/she does not. 
In this paper, we assume that each labor post their entry sheets $a$-times on average. 
In other wards, the company $k$ takes $aNP_{k}(t)$-entry sheets on average. 

We can now evaluate how many acceptances are obtained by a student and let us define the number by $s_{i}$ for each student $i\,(=1,\cdots,N)$. 
Then, we should notice that the number of acceptances for the student $i$ is defined by $s_{i}(t) = \sum_{k=1}^{K}s_{ik}(t)$ with 
\begin{equation}
s_{ik}(t) = 
\left\{
\begin{array}{cl}
\Theta(v_{k}^{*}-v_{k}(t)) \delta_{a_{ik}(t),1} & (\mbox{with prob. $1$}) \\
\Theta(v_{k}(t)-v_{k}^{*}) \delta_{a_{ik}(t),1} & (\mbox{with prob. $v_{k}^{*}/v_{k}(t)$})
\end{array}
\right.
\label{eq:def_s}
\end{equation}
 where $\Theta (\cdots)$ denotes the step function and $\delta_{a,b}$ stands for the Kronecker delta. 
 Thus, equation (\ref{eq:def_s}) means that the $s_{ik}$ takes $1$ 
 when the student $i$ posts the sheet to the company $k$ and the total number of sheets the company $k$ gathers does not exceed the quota $v_{k}^{*}$. 
 On the other hand, the variable $s_{ik}$ also takes $1$ with probability $v_{k}^{*}/v_{k}(t)$ even if 
 $v_{k}(t)>v_{k}^{*}$ holds. 
 In other words, for $v_{k}(t)>v_{k}^{*}$, the $v_{k}^{*}$ students are randomly selected as winners' from $v_{k}(t)$ candidates.

In following, we investigate statistical properties of these microscopic quantities by means of their distribution. 
\subsection{The distribution of physical quantities}
For the model introduced in the previous section, we evaluate the distribution of several microscopic physical quantities. 
To calculate the distribution numerically, we define the distribution of such quantity $A(t)$ by 
\begin{equation}
P(A)  = 
\lim_{T \to \infty} 
\frac{1}{T}
\sum_{t=0}^{T-1}
\delta_{A(t),\,A}. 
\label{eq:Pnk}
\end{equation}
Using the above definition, the distribution of the number of entry sheets $v_{k}(t)$ which the company $k\, (=1,\cdots,K)$ obtains is evaluated by substituting $A(t) = v_{k}(t)$ into the above definition (\ref{eq:Pnk}) for finite system size $K,N (\gg 1)$. 
After recursively updating the equations (\ref{eq:prob})(\ref{eq:norm}) for $T$\,($ \gg 1$)-times, 
we obtain the distribution of any micro- or macroscopic quantities through (\ref{eq:Pnk}). 

\begin{figure}[ht]
\begin{center}
\includegraphics[width=5.5cm]{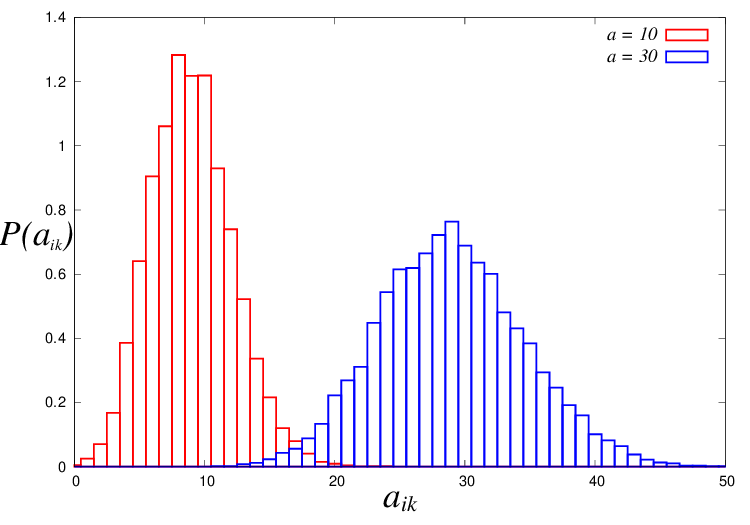}
\includegraphics[width=5.5cm]{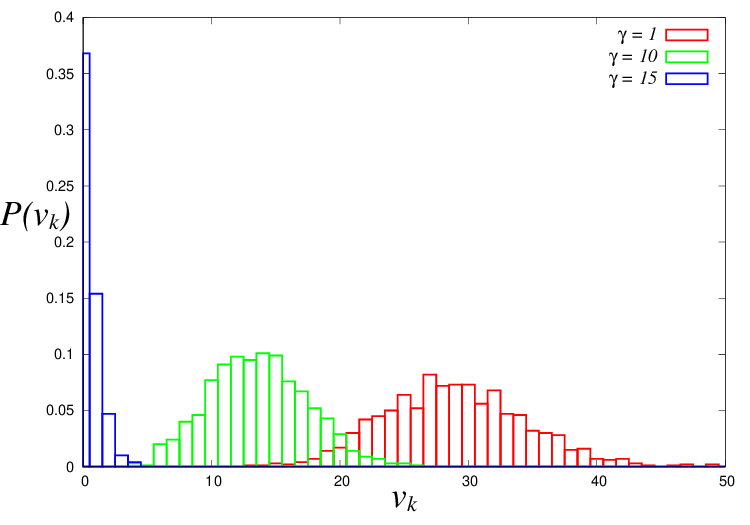}
\end{center}
\caption{\footnotesize 
The distribution of $a_{ik}$ (left). In our simulations, we set 
$N=10000, K=1000$ and for simplicity $v_{k}^{*}=v=30$, which reads $\alpha=3$. The time step to evaluate (\ref{eq:Pnk}) is set to $T=10^{4}$. 
$\beta$ and $\gamma$ are fixed as $\beta=\gamma=1$. 
We plot the $P(a_{ik})$ for $a=10$ and $30$. 
The right panel shows the distribution of the number of entry sheets the company $k$ obtains. In our simulations, we set $N=10000, K=1000$ and $v_{k}^{*}=v=30$ leading to $\alpha=3$. 
The time step to evaluate (\ref{eq:Pnk}) is set to $T=10^{4}$. 
We set $\beta=1$ and change $\gamma$ as $\gamma=1,10$ and $15$.
}
\label{fig:fg1}
\end{figure}

In Fig. \ref{fig:fg1} (left), we show the distribution of the microscopic quantity $a_{ik}$.
The system size is chosen as $N=10000, K=1000$ 
and fix the quota for each company so as to be a company-independent variable as $v_{k}^{*}=v \equiv 30$ (homogeneous quota) for simplicity. 
This choice leads to $\alpha=3$. 
The time step to evaluate (\ref{eq:Pnk}) is set to $T=10^{4}$ and the ranking factor $\gamma$ is fixed as $\gamma=1$.  
From this figure, we find that $a_{ik}$ is distributed around the average $a$. 
In our computer simulations to be given in the next sections, we use the $P(a_{ik})$ to determine the number of entry sheets posted by each student.  

We next consider the distribution of the number of entry sheets obtained by company $k$. 
We plot the result in Fig. \ref{fig:fg1} (right). 
From this figure, we clearly find that in the regime of market history actually works (the market history is switched on) $\beta=\gamma=1$, 
the distribution has a single peak around the relatively large value of $v_{k}$. However, as one increases the ranking effect $\gamma$ 
as $\gamma \gg \beta$, the peak eventually moves to zero. 
The fraction (probability) of companies getting no entry sheet is easily evaluated as follows. 

Obviously, the probability $P(v_{k} =\overline{v})$ follows a binomial distribution $P(v_{k} =\overline{v})=
{}_{aN}C_{\overline{v}}P_{k}^{\overline{v}}(1-P_{k})^{aN-\overline{v}}$, 
where we neglected the time-dependent part in $P_{k}(t)$, that is $\sim -\beta |v_{k}^{*}-v_{k}(t)|$ for $\gamma \gg \beta$. 
Then, $P(v_{k}=0)$ is written as $P(v_{k}=0) = 
(1-P_{k})^{aN} \simeq 
{\exp}(-aNP_{k})$, where $P_{k}$ is roughly estimated in the thermodynamic limit as 
\begin{eqnarray}
P_{k} & = & 
\frac{(1+k/K)^{\gamma}}{\sum_{k=1}^{K}(1+k/K)^{\gamma}} \simeq 
\frac{(1+k/K)^{\gamma}}
{\int_{1}^{K}(1+k/K)^{\gamma}dk}  =
\frac{(\gamma+1)}{K}
\frac{(1+k/K)^{\gamma}}{2^{\gamma+1}-1}.
\end{eqnarray}
Hence, in the limit of $K \propto N \to \infty$ (in other words, $K, N \to \infty$ keeping $\rho \equiv K/N ={\cal O}(1)$), 
the highest ranking company has $P_{k=K} \simeq (\gamma+1)/2K$, whereas the lowest ranking company gets 
$P_{k=1} \simeq (\gamma +1)/2^{\gamma+1}K$. 
Substituting these results into $P(v_{k}=0)$, we immediately have 
\begin{eqnarray}
P(v_{K}=0) & = & 
{\exp}[-aN(\gamma+1)/2K] 
\label{eq:Pvk0} \\
P(v_{1}=0) & = & 
{\exp}[-aN(\gamma+1)/2^{\gamma+1}K]
\label{eq:Pv10}.
\end{eqnarray}
These results imply that in the limit of $\gamma \to \infty$, the lowest ranking company gets no entry sheet with probability $P(v_{1}=0)={\exp}(0)=1$ from 
(\ref{eq:Pvk0}), whereas the highest ranking company always can get macroscopic order of entry sheet as $P(v_{k}=0) =0$ from (\ref{eq:Pv10}). 
We easily notice that the argument for (\ref{eq:Pv10}) should be valid for any companies which satisfy $k \ll K$. Therefore, for large $\gamma \gg 1$, 
macroscopic number of companies completely lose their applicants (the entry sheets) by this probabilistic nature. 
We can actually confirm this result indirectly from Fig. \ref{fig:fg1} (right) as $P(v_{k})=0$ for $\gamma=15$ which is relatively a large value in our simulations. 

In following, we investigate the macroscopic behaviour of this system through several physical quantities.
\section{Unemployment rate: Macroscopic quantity}
In the previous section, we modeled the microscopic matching process between students 
and companies by means of the probability distribution $P_{k}(t)$ of Boltzmann-Gibbs-type. 
As our main purpose is to reconstruct the macroscopic behaviour of labor markets from the microscopic description, 
we should calculate the macroscopic quantities by means of the microscopic variables. 
To reconsider the results in the previous section from the macroscopic viewpoint, here we can calculate the unemployment rate $U$ as a function of $t$ as follows. 
\begin{equation}
U_{t} =  \frac{1}{N}\sum_{i=1}^{N}
\delta_{s_{i}(t),\,0}
\end{equation}
Namely, the unemployment rate at business year $t$ is defined as a ratio of students who could not get any job ($s_{i}(t)=0$) to the total number of students $N$.
\subsection{The order parameter}
To discuss the macroscopic quantity, we consider the long time average of $U_{t}$ as an `order parameter' $U$ as usually used in statistical physics. 
Namely, we define the order parameter as follows.  
\begin{eqnarray}
U & = & 
\lim_{T\to \infty}
\frac{1}{T} \sum_{t=0}^{T-1}U_{t}
\label{eq:orderP} 
\end{eqnarray}
Here it should be noted that the above time average should be identical to the ensemble average $\langle U \rangle$, where the bracket 
$\langle \cdots \rangle$ stands for the average over the joint probability for the microscopic quantities  
$P(\mbox{\boldmath $a$}_{1}, \cdots, 
\mbox{\boldmath $a$}_{K}; 
s_{1}, \cdots,s_{N})$ with 
$\mbox{\boldmath $a$}_{k} \equiv 
(a_{1k},\cdots,a_{Nk}), 
a_{ik} \in \{0,1\}, 
s_{i} \in \{0,1,2,\cdots,a\}$, 
$i=1,\cdots, N, k=1,\cdots, K$ 
when the system can reach the equilibrium state.  
\subsection{The Beveridge curve}
\begin{figure}[ht]
\begin{center}
\includegraphics[width=5.5cm]{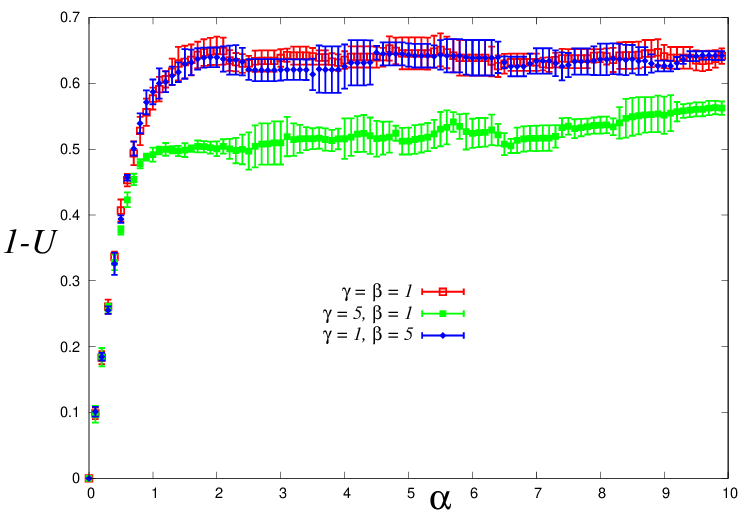}
\includegraphics[width=5.5cm]{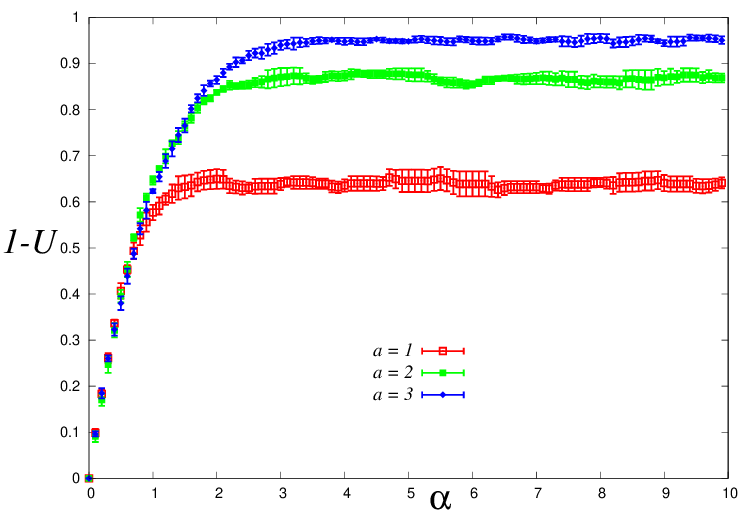}
\end{center}
\caption{\footnotesize 
The employment rate $1-U$ as a function of $\alpha$ (left). 
We evaluated the rate for the cases $(\gamma, \beta)=(1,1),(1,5),(5,1)$.  
In our simulations, we set the system sizes as $N=500, K=50$. The errorbars are calculated by five independent trials.
The right panel shows the employment rate $1-U$ as a function of $\alpha$ for the case of $a=1,2$ and $a=3$ keeping $\gamma=\beta=1$. 
}
\label{fig:fg4}
\end{figure}

In Fig. \ref{fig:fg4} (left), we plot the employment rate $1-U$ as a function of 
$\alpha$ for several choices of $(\gamma, \beta)=(1,1),(1,5),(5,1)$.  
From this figure, we certainly find that the employment rate is lower than $\sim 0.7$. 
In these numerical simulations, we assumed that each student posts their entry sheet just only once on average. 
However, if the number of posting increases the situation might be changed. 
Thus, we next consider the case in which each student posts their applications $a$ times on average. 
Therefore, we increases the number of entry sheets $a$ for each student to post to the market. 
We check the case $a=1,2$ and $a=3$. 
The result is shown in Fig. \ref{fig:fg4} (right). 
From this figure, as we expected, we find that the employment rate increases up to near $1$ (perfect employment) when we increase the number $a$. 

We should notice that the plots shown in Fig. \ref{fig:fg4}(left) and 
Fig. \ref{fig:fg4} (right) correspond to the so-called 
{\it Beveridge curve} in economics (labor science). 
Usually, the Beveridge curve is defined as the behaviour of the number of vacancy $V$ against the unemployment rate $U$ (hence, it is sometimes referred to as {\it UV curve}). 
However, as we already mentioned, in the thermodynamic limit $V,N \to \infty$, the relevant system parameter is job offer rate $\alpha$ rather than the number of vacancies $V$. Thus, we might regard the $\alpha$ as the effective number of vacancies. 
Hence, the $U$-$\alpha$ curves shown in these figures correspond to the conventional Beveridge curves. 
\section{Phase transitions in labor markets}
\begin{figure}[ht]
\begin{center}
\includegraphics[width=5.8cm]{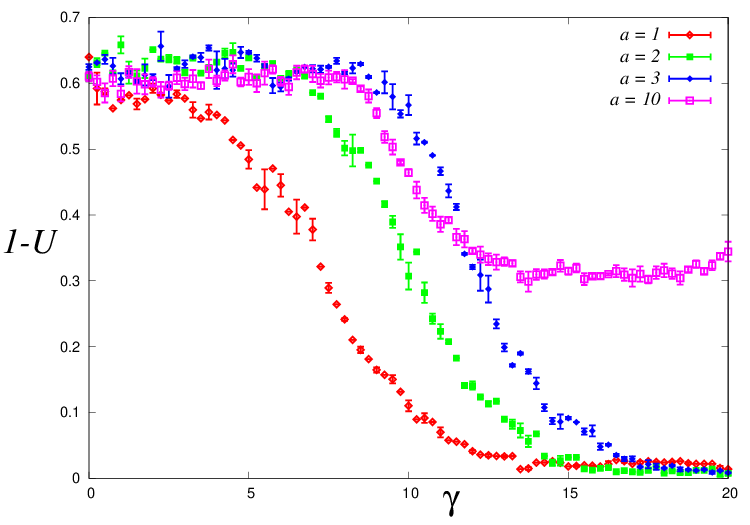}
\includegraphics[width=5.8cm]{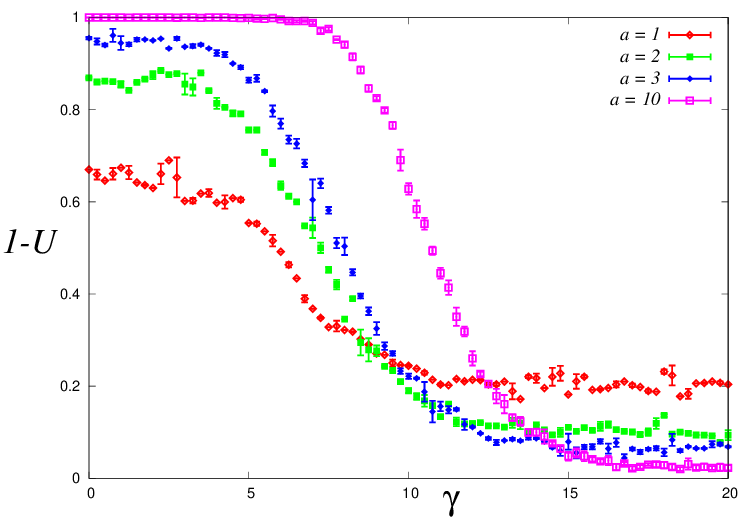}
\end{center}
\caption{\footnotesize 
The employment rate $1-U$ as a function of 
$\gamma$ for a fixed $\beta$ value ($\beta=1$). 
The results are plotted for $a=1,2,3$ and $a=10$. 
The left panel is given for $\alpha=1$, whereas the right panel is obtained for $\alpha=10$.
}
\label{fig:fg6}
\end{figure}

We next consider the $\gamma$-dependence of the employment rate for several values of $a$. 
In Fig. \ref{fig:fg6}, we show the employment rate $1-U$ as a function of 
$\gamma$ for a fixed $\beta$ value ($\beta=1$). 
The results are plotted for $a=1,2,3$ and $a=10$. 
The left panel is given for $\alpha=1$, whereas the right panel is obtained for $\alpha=10$. From this panel, as we expected, we find that the employment rate for a relatively high job offer ratio $\alpha=10$ increases up to near $1$ when we increase the number $a$. 

One of the remarkable features of the results is existing a sort of `phase transitions' in our probabilistic labor market. 
Namely, in the lower panel in Fig. \ref{fig:fg6}, 
we clearly find that there exist two distinct phases, namely, `perfect employment phase' ($1-U \simeq 1$) and `perfect unemployment phase' ($1-U \simeq 0$, 
that is, $U \simeq 1$), and the system changes gradually from the perfect employment phase to the poor employment phase around $\gamma \simeq 10 (\equiv \gamma_{c})$. 
To evaluate the residual employment rate, we consider the extreme limit $\gamma \to \infty$ for the simplest case $\alpha=a=1$. 
In this limit, all students want to post their own single entry sheet to the highest-ranking company $k=K$. 
As the result, only $v_{K}^{*}=v=10$ students get their jobs and the residual employment rate is approximately evaluated as $1-U \simeq v/N=10/500=0.02$.

From the viewpoint of labor markets, the phase transition might be understood as follows. 
For the case of a society with quite high job offer ratio such as $\alpha=10$, the employment rate can be almost 
$1$ when each labors can post $a=10$ entry sheets on almost randomly to companies. 
However, as labors start to take into account the ranking of company, namely, for $\gamma \simeq \gamma_{c}$, the employment rate is gradually decreased to the zero-level. 
Apparently, this result is induced due to the global mismatch between students and companies as observed in recent Japanese labor market for university graduates.  
\section{The Philips curve}
In the previous section, we have made a simple probabilistic model for job matching process between university graduates and companies in Japan. 
We evaluated the unemployment rate $U$ as an order parameter and found that the system undergoes a phase transition when one changes the system parameters. 

We next consider the Philips curve for our labor market. 
To obtain the Philips curve, we should evaluate the inflation rate separately 
and it needs some information about the production process of the companies and consumption procedure by consumers. 
Moreover, the bargaining process of wages of labors for each company also should be taken into account to determine the inflation rate. 
In this paper, we shall use the macroscopic formula for the inflation rate given by Neugart \cite{Neugart}. 
Then, by making coupled equations with our result of unemployment rate, we attempt to draw the Philips curve. 
\subsection{Macroscopic Neugart model}
In this subsection, we shall briefly explain the derivation of non-linear maps with respect to the unemployment 
and inflation rates according to Neugart \cite{Neugart}. 

We first defined the update rule for the unemployment rate as follows. 
\begin{equation}
U_{t+1}  =  
U_{t} + \xi (1-U_{t}) -o_{t}U_{t}
\label{eq:update_U}
\end{equation}
where the second term in the right hand side denotes the contribution of labors who are fired (lost) their jobs at time $t$ and the ratio is controlled by a single parameter $\xi$. 
The third term means the contribution of labors who get their job at time 
$t$ and $o_{t}$ is a time-dependent rate. 
The $o_{t}$ is explicitly given by 
\begin{equation}
o_{t} = 
\frac{J_{s} + \Gamma (m-\pi_{t})}
{U_{t}+d(1-U_{t})}
\label{eq:def_o}
\end{equation}
where the denominator denotes the total amount of labors who seek for the jobs at time $t$ and 
$d(1-U_{t})$ corresponds to the labors who are `on the job searching'. 
On the other hand, in the numerator, 
$J_{s}$ denotes the time-independent number of job openings  and the time-dependent part of job openings comes from the second term 
$\Gamma (m-\pi_{t})$. 
In this term,  $\pi_{t}$ denotes the inflation rate at time $t$ and 
$m$ stands for a constant growth rate for the value of money. 
Hence, equation (\ref{eq:def_o}) means that the probability getting jobs $o_{t}$ decreases when the number of labors who seek for jobs increases, 
and increases when the inflation rate is smaller than the growth rate for the value of money. 

On the other hand, the inflation rate at time $t$ is written in terms of the `expected inflation rate' $\pi_{e,t}$ as follows. 
\begin{equation}
\pi_{t}  =  
\frac{1}{\delta}
\left(
\pi_{e,t}+ 
\frac{w_{b,t}-w_{p}}{w_{p}}
\right)=
\frac{1}{\delta}
\left(
\pi_{e,t} + 
\frac{\mu-(1-c_{2})U_{t}}
{1-\mu}
\right)
\label{eq:def_Epi}
\end{equation}
where $\delta$ stands for a scaling factor and $w_{b,t}$ is bargaining wages and we naturally set $w_{b,t}=1-(1-c_{2})U_{t}$ with a constant $0 \leq c_{2} \leq 1$. 
The justification of this choice depends on the validity of our assumption that the bargaining should go well when the unemployment rate is low. 
The union having enough number of labors can negotiate with management for wage increases well.
The $w_{p}$ is a base wage and it is controlled a single parameter $\mu$ as 
$w_{p}=1-\mu$. 

Then, the expected inflation rate is updated by means of a linear combination of inflation rate 
$\pi_{t}$ and the expected inflation rate $\pi_{e,t}$ with $0 \leq c_{1} \leq 1$ as follows. 
\begin{equation}
\pi_{e,t+1} = 
c_{1} \pi_{t}+ 
(1-c_{1}) \pi_{e,t}
\label{eq:update_pi}
\end{equation}
From equations (\ref{eq:update_U})(\ref{eq:def_o})(\ref{eq:def_Epi}) and 
(\ref{eq:update_pi}), we obtain the following non-linear maps with respect to $U$ and $\pi$: 
\begin{eqnarray}
U_{t+1}  & = &  U_{t}+\xi (1-U_{t})-U_{t}\frac{J_{s}+\Gamma(m-\pi_{t})}{U_{t}+d(1-U_{t})} 
\label{ut} \\
\pi_{t+1}  & = & \frac{1}{\delta} 
{\biggr (}   \frac{\mu}{1-\mu}+c_{1}\pi_{t} + (1-c_{1}) 
{\biggr (} \delta\pi_{t} -  \frac{\mu -(1-c_{2})U_{t}}{1-\mu}{\biggr )}
{\biggr )} \nonumber \\
\mbox{} & - & \frac{1}{\delta}
{\biggr (} 
\frac{1-c_{2}}{1-\mu} {\biggr (} U_{t}+\xi (1-U_{t}) -  U_{t}\frac{J_{s}+\Gamma(m-\pi_{t})}{U_{t}+d(1-U_{t})}
{\biggr )}
{\biggr )} 
\label{eq:pi}
\end{eqnarray}
The fixed point of the above non-linear maps is easily obtained as 
$(U^{*},\pi^{*}) = 
(\{\mu-m(\delta -1)(1-\mu)\}/(1-b),m)$. 
Then, according to 
Neugart \cite{Neugart}, we set the value of $J_{s}$ in terms of the above fixed point, namely, 
by inserting the fixed point $U_{t+1}=U_{t}=U^{*}, \pi_{t+1}=\pi_{t}=\pi^{*}$ into the equation (\ref{ut}), we obtain 
$J_{s}=J_{s}^{*} \equiv \xi(1-U_{*})(U^{*}+d(1-U^{*}))/U^{*}$. 

\begin{figure}[ht]
\begin{center}
\includegraphics[width=7cm]{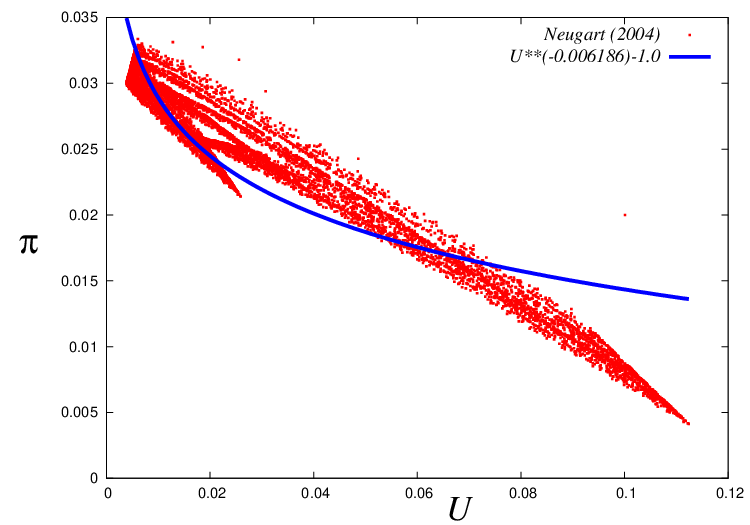}
\end{center}
\caption{\footnotesize 
Philips curve obtained by the Neugart model \cite{Neugart}.
}
\label{fig:fg9}
\end{figure}

The chaotic attractor $(U_{t},\pi_{y})$ gives the Philips curve. 
In Fig. \ref{fig:fg9}, we plot the Philips curve obtained by the set of parameters: 
$\xi=0.18, 
d=0.01,c_{1}=c_{2}=0.5, 
\mu=0.04, 
\Gamma=0.5, \delta =2$ and 
$m=0.03$. 
From this figure, we observe that the chaotic attractor follows the scaling form: $\pi + 1.0 \propto U^{-0.006186}$. 
\subsection{Coupling with our probabilistic model}
As is shown in the previous sections, 
we can make a model for the probabilistic labor market which described by $a_{ik}(t)$: (\ref{eq:aik}) and $s_{i}(t)$: (\ref{eq:def_s}) with $s_{i}(t)=\sum_{k}s_{ik}(t)$ microscopically. 
The behaviour of the system is macroscopically written in terms of the time-dependence of unemployment rate 
$U_{t}$ or the order parameter $U$ which is defined as long-time average of the $U_{t}$ as 
$U=\lim_{T \to \infty} (1/T)\sum_{t=0}^{T-1}U_{t}$. 
To draw the Philips curve, here we consider the coupled equations for $U_{t}$ obtained in our model and $\pi_{t}$ which is one of the no-linear map in the Neugart model. 
Namely, here we use the equation (\ref{eq:pi}) with $J_{s}=J_{s}^{*}$ for the update rule for the inflation rate. The other parameters in the Neugart model are set to 
$\xi=0.18, 
d=0.01,c_{1}=c_{2}=0.5, 
\mu=0.04, 
\Gamma=0.5, \delta =2$ and 
$m=0.03$. 
\subsubsection{Typical dynamics}
\begin{figure}[ht]
\begin{center}
\includegraphics[width=5.5cm]{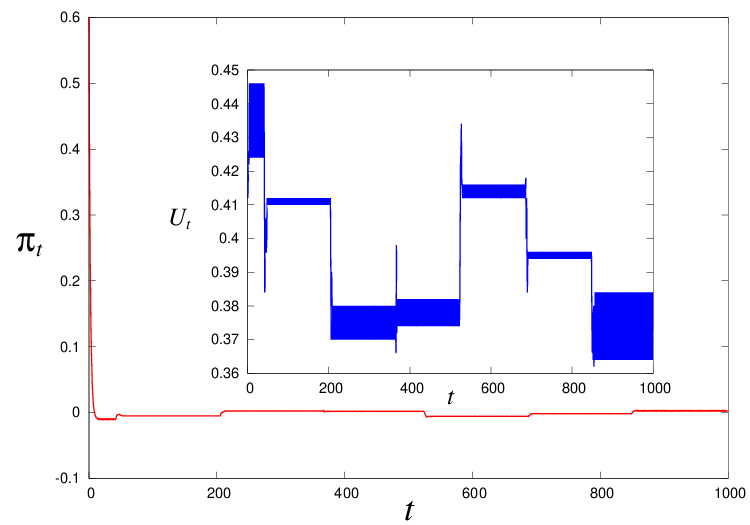}
\includegraphics[width=5.5cm]{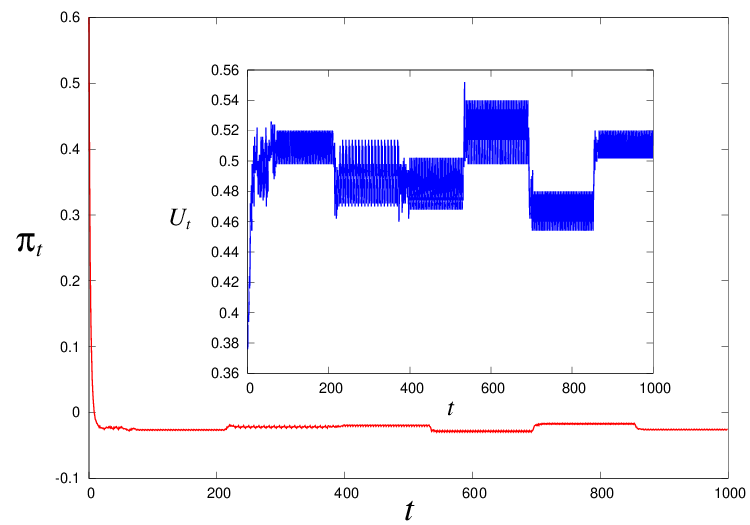}
\end{center}
\caption{\footnotesize 
Typical dynamics of the 
unemployment rate $U_{t}$ and 
inflation rate $\pi_{t}$. 
The upper panel is given for $\beta=\gamma=1$, 
whereas the lower panel is plotted for 
$\beta=10,\gamma=1$. 
We set $N=500,K=50, v_{k}^{*}=10 \,(\alpha=1)$ and $a=10$. 
We find that 
both $U_{t}$ and $\pi_{t}$ are 
`clustered' during the interval $ 100 \sim 200$. 
Within each interval of `clustering', the these quantities behave periodically (oscillate). }
\label{fig:fg09}
\end{figure}

We first show the typical dynamics of the 
unemployment rate $U_{t}$ and 
inflation rate $\pi_{t}$ in Fig. \ref{fig:fg09}. 
From this figure we find that 
both $U_{t}$ and $\pi_{t}$ are 
`clustered' during the interval $ 100 \sim 200$. 
Within each interval, the these quantities behave periodically (oscillate).  @ 
\mbox{}

\begin{figure}[ht]
\begin{center}
\includegraphics[width=7cm]{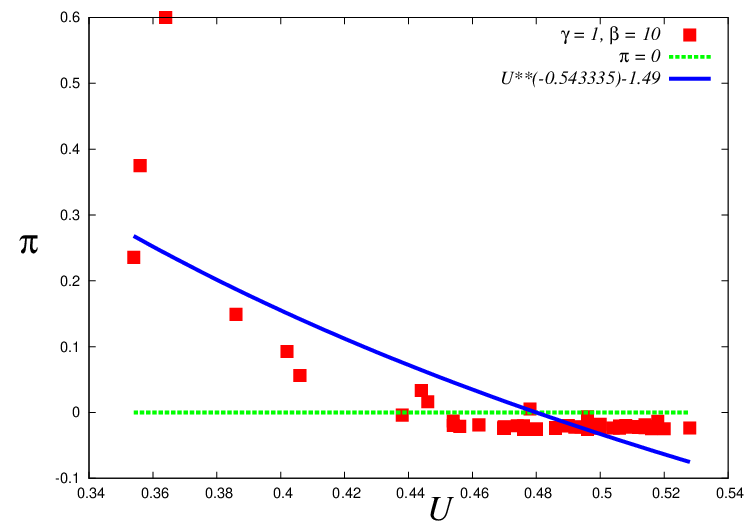}
\end{center}
\caption{\footnotesize 
The Philips curve for our probabilistic model for $\beta=10,\gamma=1$. 
We set $N=500,K=50, v=10 \,(\alpha=1)$ and $a=10$.
The curve is `well-fitted' the form: $\pi +1.49 \propto U^{-0.54}$. 
 }
\label{fig:fg099}
\end{figure}

Finally we plot the Philips curve 
as a trajectory $(U_{t},\pi_{t})$ of the dynamics. 
The results are shown in Fig. \ref{fig:fg099}. We find that the negative correlation 
between $U_t$ and $\pi_{t}$ is actually observed for $\beta=10,\gamma=1$ and  the curve is `well-fitted' by the form: $\pi +1.49 \propto U^{-0.54}$. 
\section{Summary}
In this paper, on the basis of statistical physics, we proposed a minimal model to describe Japanese labor markets from the microscopic point of view. 
The model is definitely the simplest one and it might be possible for us to consider various extensions.
\section*{Acknowledgment}
We thank organizers of {\it Econophysics-Kolkata VI}, in particular, Frederic Abergel, Anirban Chakraborti, 
Asim K. Ghosh and Bikas K. Chakrabarti.  
We also thank Enrico Scalas, Giacomo Livan, Nobuyasu Ito, Koji Oishi, Takero Ibuki and Yu Chen  for valuable discussion. 
This work was financially supported by 
Grant-in-Aid for Scientific Research (C) of Japan Society for the Promotion of Science, No. 22500195. 

\end{document}